# Electrons on a straight path: A novel ionisation vacuum gauge suitable as reference standard

Karl Jousten[1], Matthias Bernien[1], Frédéric Boineau[2], Nenad Bundaleski[3], Claus Illgen[1], Berthold Jenninger[4], Gustav Jönsson[5], Janez Šetina[6], Orlando M.N.D. Teodoro[3], Martin Vičar[7]

[1] Physikalisch-Technische Bundesanstalt (PTB), Abbestr. 2-12, 10587 Berlin, Germany,
[2] Laboratoire national de métrologie et d'essais (LNE), 1 rue Gaston Boissier, 75724 Paris Cedex 15, France
[3] CEFITEC, Department of Physics, Faculty of Sciences and Technology, Nova University of Lisbon 2829-515 Caparica, Portugal
[4] CERN European Organization for Nuclear Research, 1211 Geneva 23 Switzerland
[5] RISE Research Institutes of Sweden AB, Brinellgatan 4, BOX 857, SE-501 15 Borås, Sweden
[6] IMT Institute of Metals and Technology, Lepi pot 11, 1000 Ljubljana, Slovenia
[7] Czech Metrology Institute (CMI), Okruzni 772/31, 638 00 Brno, Czech Republic

Corresponding author: karl.jousten@ptb.de

## Abstract

The consortium of the European project 16NRM05 designed a novel ionisation vacuum gauge in which the electrons take a straight path from the emitting cathode through the ionisation space into a Faraday cup. Compared to existing ionisation vacuum gauges, this has the advantage that the electron path length is well defined. It is independent of the point and angle of emission and is not affected by space charge around the collector. In addition, the electrons do not hit the anode where they can be reflected, generate secondary electrons or cause desorption of neutrals or ions. This design was chosen in order to develop a more stable ionisation vacuum gauge suitable as reference standard in the range of $10^{-6}$ Pa to $10^{-2}$ Pa for calibration purposes of other vacuum gauges and quadrupole mass spectrometers. Prototype gauges were produced by two different manufacturers and showed predictable sensitivities with a very small spread (< 1.5%), very good short-term repeatability (< 0.05%) and reproducibility (< 1%), even after changing the emission cathode and drop-down tests. These characteristics make the gauge also attractive for industrial applications, because a gauge exchange does not require calibration or re-adjustment of a process.

1. Introduction

The ionisation vacuum gauge [1] is the measuring instrument for high and ultrahigh vacuum. While for many applications only a modest measurement accuracy is required, other applications like the measuring of pumping speeds of high vacuum pumps, calibrations and comparisons between national standards would greatly benefit from a good accuracy. This should not only cover nitrogen gas, but all gas species relevant for applications in high and ultrahigh vacuum, i.e. relative gas sensitivity factors [1] should be known with small uncertainties. The latter feature is of great value in industrial applications, where the true pressure for other gas species than nitrogen needs to be known.

Presently available commercial ionisation vacuum gauges, mostly of the Bayard-Alpert (BA) type, lack of measurement accuracy, robustness and reproducibility for the mentioned applications [2]. Some authors and manufacturers have significantly improved BA gauges [3], but not to the extent that this type of ionisation vacuum gauge meets the requirements for the applications mentioned above, in particular not for accuracy of relative gas sensitivity factors. Since the latter depends on the type of gauge with a big spread between the individual gauges, the standard committee ISO TC 112 made a statement for the need of research on a stable and accurate ionisation vacuum gauge that can be standardised.

The European project EMPIR 16NRM05 joining 5 National Metrology Institutes, CERN, the Nova University of Lisbon and two gauge manufacturers carried out such research. A literature review [2] made clear that it will not be possible to achieve the goal of an ionisation vacuum gauge with stable sensitivity and stable relative gas sensitivity factors with the most widespread BA-design. The following reasons were identified:

1. The instability of the electron emission distribution from the cathode and variance of electron path lengths and energy

A changing distribution causes different electron paths and lengths and a changing sensitivity. Changes of the emission distribution are caused by varying cathode temperatures and potentials, as well as work function changes due to contamination and ion bombardment. Changes in the electron emission are also caused by geometrical variation and instability of the cathode in most of today's commercial gauges. Due to the undefined trajectory in the BA gauge there is also a broad variance of electron energy inside the anode.

2. Lack of mechanical rigidness and reproducibility

Cathode, anode and collector in the BA design are made of wires. The collector wire may be as thin as 10 µm. It is very difficult to manufacture and position the electrodes in a reproducible manner. In addition, the electrodes are prone to be deformed by shocks or vibration during transport. A cathode exchange requires a re-calibration of the BA gauge. These reasons lead to transport instability and a large scatter of sensitivities due to different electron trajectories in each gauge.

3. Space charge effects

Positive space charge around the collector, in particular at higher pressure, perturbs the field distribution and affects ion and electron trajectories. The space charge depends on the gas species and the number of secondary electrons and is usually not stable.

4. Electron stimulated desorption (ESD) of neutrals and ions from the anode and backscattering on the anode

The number of desorbed ions and molecules changes with time due to the changes of the surface condition and can make a significant contribution to the ion current compared to the one by gas phase ions. When electrons hit the anode grid, they may be backscattered, thus complicating their trajectories.

5. Soft X-rays produced by electrons impinging on the anode.

The soft X-rays generate numerous electrons within the gauge which cannot be controlled and will change the measured ion current in numerous ways.

6. Secondary electrons produced on the collector by ion impingement.

The secondary electron yield on the collector will depend on its surface condition, which is changing with time.

Another popular design type, the extractor, reduces the effects of electron stimulated ion desorption (ESD) and X-rays [4], but the other effects, in particular the effect of instability of the electron emission distribution from the cathode, are still significant.

For this reason, for the purpose of a more stable ionisation gauge, we developed a design [5] based on ideas of Bills et al. [6] and Klopfer [7]. These designs offer both the possibility of a well-defined electron path and the possibility to separate at least some of the surface effects from volume effects. The Klopfer design, which we finally selected, also offers the possibility to use a kind of point emitter of electrons avoiding the problem of locally changing electron emission.

Present computer capabilities allow to simulate electron and ion trajectories in an ionisation vacuum gauge within a reasonable amount of time. For this reason, we developed the design completely by simulation using the software packages OPERA and SIMION, and at a later stage also COMSOL. The design will be detailed in the next section. In Section 3 we will summarise our studies on the ion induced secondary electron yield from the collector, the only of the mentioned effects which we could not avoid in our design. Section 4 will report on the test results obtained with prototype gauges produced by two different manufacturers, followed by a concluding section.

**2. Design and simulation**

Cathodes with thermionic emission are still more reliable than the ones with cold emission. To address the Problem 1 of instable emission with such cathodes, we took two measures:

a. We use thermionic emission from a disk attached to a heating wire.

b. We use a Wehnelt-electrode around it.

A Wehnelt electrode is a cylindrical electrode around a thermionic emission cathode with an aperture to an anode and with such a negative bias relative to the cathode that a small region of the cathode has a net electric field (due to both anode attraction through the aperture and Wehnelt repulsion) that allows electron emission from only that area of the cathode.

With this assembly, all emitted electrons are guided forward and enter the ionisation volume in a rather parallel beam (Figure 1, top). No matter from which location on the emitter disk the electrons leave, they follow closely the same trajectories inside the ionisation volume. The electron path length is not affected by local changes of work-functions and temperatures.

To further address Problem 1, we carefully guided the electron beam by the anode and the ion collector and defined the region where the electrons have enough energy to ionise molecules and where these generated ions are drawn to the collector. The cylindrical anode is provided with two apertures at the entry and exit of the electrons into and out of the anode cage. The entry and exit plane of the anode cage defines the boundaries. at which ions are collected by the ion collector. This length of the electron beam of 50 mm between the two planes (Figure 1, bottom left) is the nominal electron path length and is used to calculate the ion collector current and sensitivity of the gauge.

The ion collector is composed of two elements, a focussing ring and a rod (Figure 1). The collector ring has the function of central electrode of an electrostatic lens which focuses the electron beam through the cylindrical anode into the exit aperture. In this way, emitter misalignments of more than 0.3 mm can be tolerated in all directions. The collector rod extracts the ions from the electron beam over its full length in the anode cylinder by applying a small extraction potential to overcome the space charge potential inside of the electron beam. This also results in a small deflection of the electron beam inside the ionisation volume.

After exiting the ionisation chamber, a deflector directs the electrons into a Faraday cup. The ratio of electron current arriving in the Faraday cup and the emitted current leaving the emitter defines the electron transmission efficiency $T_e$.

Electrons hit the Faraday cup in an area where there is no direct view to the ion collector. Ions emitted from the Faraday cup due to the ESD effect will be efficiently collected by the deflector (Problem 4). Generated soft X-ray photons cannot directly reach the collector, which addresses Problem 5. This absence of soft X-ray photons gave us freedom in the design of the collector electrode. With the large collector surface compared to Bayard-Alpert gauges we ensure a low ion density around the collector and avoid ion space charge around the collector at high pressures eliminating Problem 3. The shape of the collector eliminates any loss of ions due to their angular momentum.

With both ion collection efficiency and electron transmission efficiency approaching 1, the sensitivity of the gauge can be determined from the ionisation path length $L$ and the energy dependent ionisation cross section $\sigma$ using the equation of an idealised ionisation vacuum gauge

$$S_{ideal} = \frac{\sigma L}{kT} \tag{1}$$

with $k$ being the Boltzmann constant and $T$ the absolute temperature. For $L = 0.05$ m and a nominal ionisation energy of 200 eV, corresponding to an ionisation cross-section of $2.27 \times 10^{-20}$ m$^2$ for $N_2$ [8] the sensitivity is 0.281 Pa$^{-1}$ for the idealised gauge. The collector ring creates a potential dip along the electron path, where the ionisation cross section is slightly enlarged. Simulations can take these variations into account, yielding sensitivities of 0.281 Pa$^{-1}$ (COMSOL), 0.288 Pa$^{-1}$ (OPERA) and 0.297 Pa$^{-1}$ (SIMION), respectively.

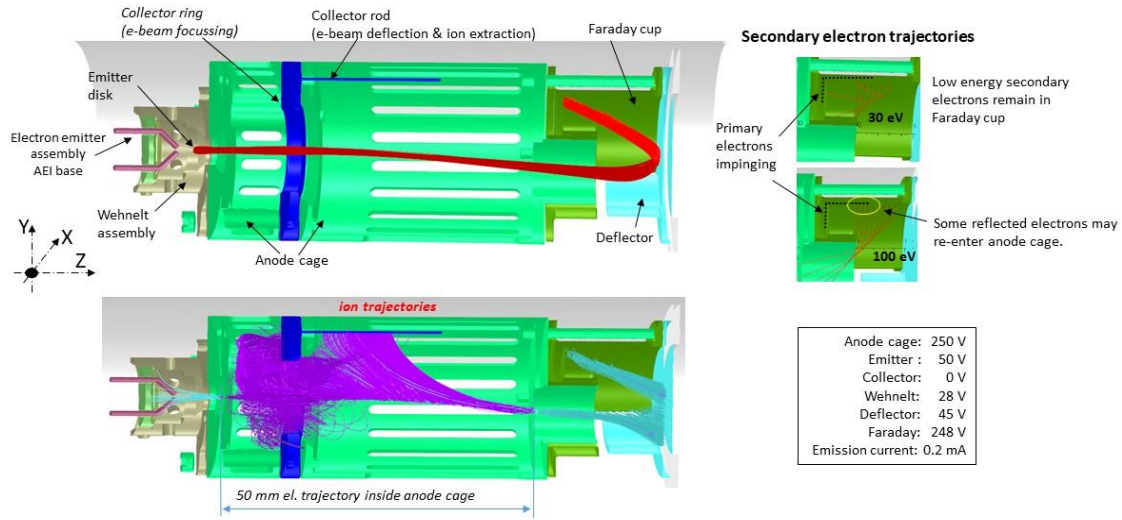

Figure 1 *Design and charged particle trajectories of the new gauge [5]. Top left: the electrode arrangement and electron trajectories. Bottom left: ion trajectories. The purple trajectories correspond to those that are collected by the ion collector. The light blue ones are those that are rejected. Top right: Trajectories of secondary electrons leaving the surface perpendicular with the indicated kinetic energy. Bottom right: the potentials and emission current applied for this simulation.*

Ions generated behind the exit plane of the ionisation volume are attracted to the deflector (Figure 1, bottom left). Furthermore, most of the secondary electrons that are generated inside the Faraday cup are pushed back by the deflector (Figure 1, bottom right), which addresses Problem 4. Some of those having higher energies may hit the small exit tube of the anode cage. This results in an apparent reduction of the electron transmission without affecting the sensitivity of the gauge, which is confirmed by measurements (see Section 4). Only an insignificant part of secondary electrons can reach the ionisation volume, also addressing Problem 4.

We also tested the robustness of the design by simulations. The different electrodes in the simulation prototypes were displaced and the range of displacements in which the performance parameters (electron transmission, collection efficiency and sensitivity) remain within the target precision of 1% was determined. Varying the emitter potential, we consider the energy distribution of the emitted electrons and account for the fact that the variation of heating currents causes variations of the emitter potential due the change in electrical potential drop along the heating wire.

One interesting aspect about the concept of electron transmission is the fact that any major disturbance or electrode misalignment will lead to a reduction of the electron transmission. On the other hand, a high transmission efficiency of $\geq 95\%$ is an indicator that there is no major disturbance, and the measurement can be trusted.

## 3. Ion induced secondary electron yield on some collector materials

Problem 6 mentioned in the introduction could not be eliminated in our design. Ions impinging on the collector will generate secondary electrons and the measured collector current will be equal to the true ion current multiplied by a factor $(1 + \gamma_{e,ion})$, where $\gamma_{e,ion}$ is the ion induced secondary electron yield (IISEY) defined as the number of emitted electrons per incident ion. The processes responsible for the electron emission in the ion energy range of 50 eV - 250 eV, explained in [2], take place within the ion penetration depth, i.e. they are restricted to the first several nanometer of the material surface. Any change of this surface will directly affect $\gamma_{e,ion}$ i.e. the gauge sensitivity, thus introducing measurement instabilities. The relative change $\Delta S$ of gauge sensitivity $S$ caused by a change of $\gamma_{e,ion}$ will be practically equal to ( $\gamma_{e,ion} \ll 1$):

$$\left(\frac{\Delta S}{S}\right)_{\gamma_{e,ion}} = \frac{\Delta \gamma_{e,ion}}{1+\gamma_{e,ion}} \quad (2)$$

Since we could not eliminate IISEY, we aimed at stabilizing $\gamma_{e,ion}$. Bombardment of the collector by $Ar^+$ ions was considered as a method to stabilise the ionisation gauge sensitivity by cleaning the collector [9]. Our recent experimental studies, however, showed that the reason for the stabilisation effect is exactly the opposite: when the partial pressure of hydrocarbons inside the gauge is sufficiently high, ion bombardment promotes hydrocarbon layer growth [10], so that the sensitivity stabilisation is achieved by reaching a sufficiently thick hydrocarbon overlayer on the collector surface. Under such conditions, IISEY becomes independent of the collector material.

Using $Ar^+$ ion bombardment for the gauge preconditioning in order to stabilize the sensitivity is expected to be successful in the case of inert gases, nitrogen and probably hydrocarbons. However, its effectiveness is questionable in the case of more reactive gases (e.g. $O_2$ or halogens) which could poison the hydrocarbon overlayer and modify the gauge sensitivity. In that respect, more investigations are necessary to evaluate the true potential of this preconditioning approach. Other approaches for stabilisation could be to prepare collector surfaces with specific chemical (highly inert) and/or morphological (highly corrugated in order to reduce $\gamma_{e,ion}$) properties. Finally, adding a suppressor grid above the collector can be another solution for a stable ($\gamma_{e,ion} \approx 0$) sensitivity. This, however, would lead to a further complication of the design and also cause other problems like the reduction of the ion collection efficiency and the acceleration of secondary electrons from the suppressor grid towards the collector.

## 4. Experimental results

Two different vacuum gauge manufacturers produced 22 prototype gauges (Figure 2) according to the design. All electrodes were made of low carbon stainless steel (ANSI: 316L, DIN 17440: 1.4404). For an emission cathode, an indirectly heated Ta disk on an AEI base, a mounting base for cathodes in electron microscopes, was used.

Commercial devices for power supplies of all electrodes including the cathode and for the measurement of electron and ion current were used. All meters were calibrated in our institutes and are traceable to primary standards. The experimental testing work reported below was shared between the laboratories who received different copies of the prototype gauges.

Before testing, the potentials of Wehnelt electrode, Faraday cup and deflector were optimised. The potentials of cathode (50 V) and anode (250 V) are crucial to the sensitivity because of the energy dependence of ionization cross section $\sigma$ in Eq. (1) and to the focusing of the electron beam, and were therefore fixed.

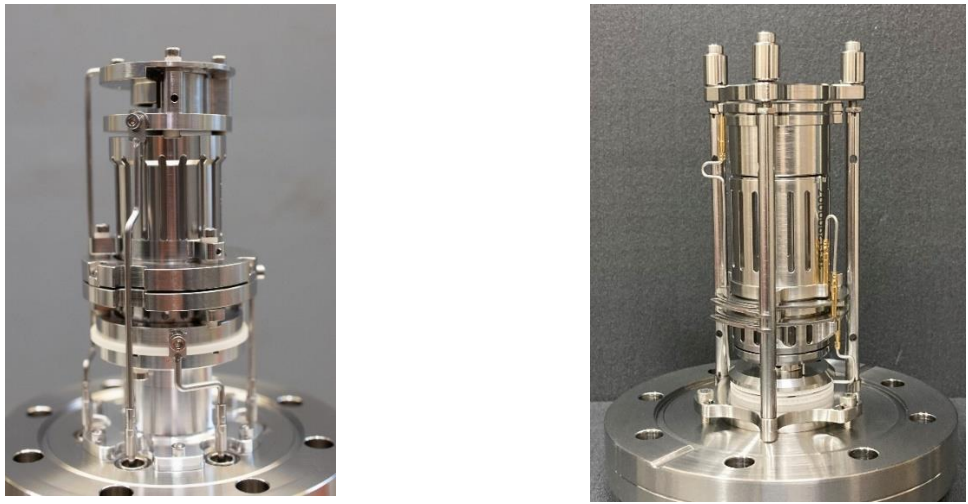

Figure 2 *Photographs of two prototype gauge types realising our design, each from a different manufacturer (INFICON and VACOM)*

It was found that the apparent electron transmission efficiency $T_e$ is 97 % for the nominal values of electrode potentials used for simulation. $T_e$ is defined as the electron current measured on the Faraday cup divided by the total emission current and may be affected by secondary electrons. An increase of Faraday potential from 248 V to 252 V increases $T_e$ to approximately 98%. This is attributed to the suppression of low energy secondary electrons from the Faraday cup. A further increase of the Faraday cup potential to 350 V lets $T_e$ increase to 99%, which is attributed to the additional suppression of high energy electrons elastically scattered from the Faraday cup. Based on these findings, a Faraday potential of 280 V was selected for further measurements, for which $T \approx 98\ \%$.

The dependence of $T_e$ on the Wehnelt potential is shown in Figure 3. For well aligned cathodes there is practically no dependence between 15 V and 35 V. For misaligned cathodes, however, a significant dependence of $T_e$ on the Wehnelt potential is observed. The sensitivity, however, remains unchanged. Figure 3 shows that the electron current reaching the Faraday increases monotonically with the Wehnelt potential. Since this current determines the ionising current, the optimal Wehnelt potential is the one, at which the emission current is maximal, and $T_e$ is still at its maximum level (34 V in Figure 3). When $T_e$ is reduced at higher Wehnelt potentials, a defocusing of the electron beam occurs leading to the unwanted effect of electrons hitting the anode causing detrimental effects by electron induced emissions.

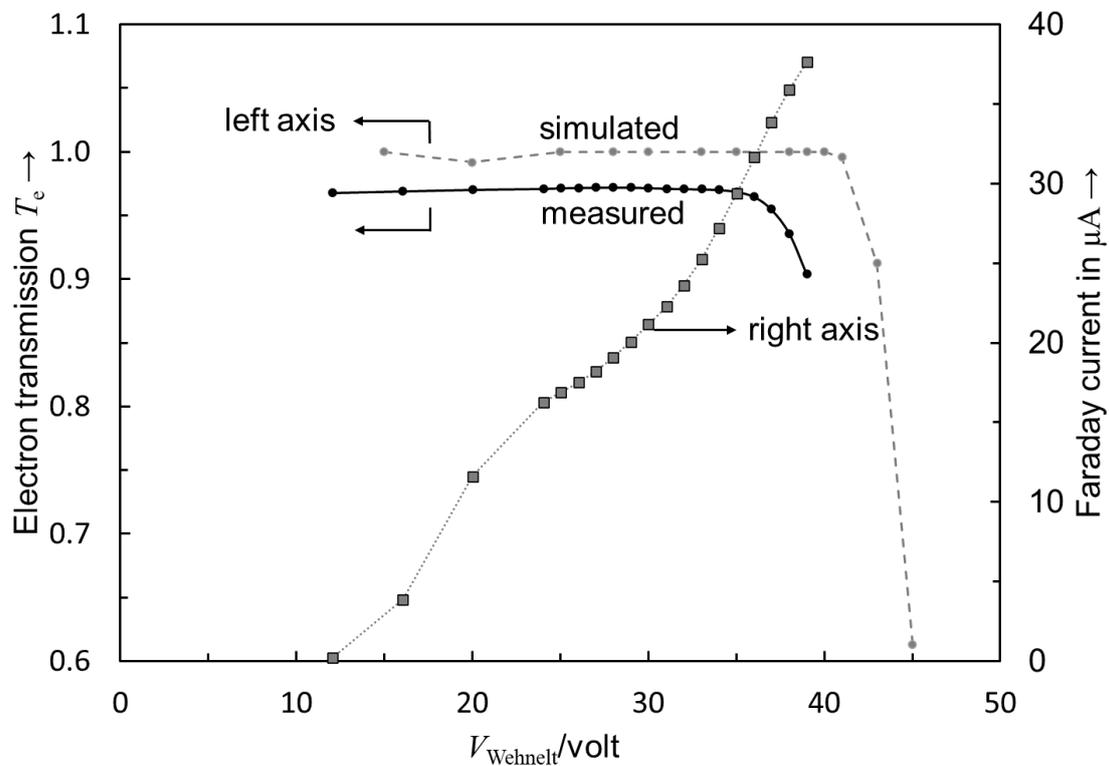

Figure 3 *Electron transmission efficiency $T_e$ simulated by OPERA and measured, grey/black circles and left axis, and measured Faraday current, grey quadrats, right axis, in dependence of the Wehnelt potential*

The main goal of the project was to achieve a design where relative gas sensitivity factors are independent of the individual gauge. The tests revealed that even absolute sensitivities, determined by comparison with vacuum primary standards, are independent of the individual gauge. The values of measured nitrogen sensitivities for different manufactured prototype gauges are presented in Figure 4, which also contains sensitivity values obtained by simulations.

The very good agreement of simulation and calibration is partly due to a coincidence: On the one hand, the simulations did not include the IISEY from the ion collector surface, on the other hand the temperature of the molecules inside the gauge head is not near 293 K to 300 K as assumed in the simulations. The investigations of IISEY indicate that the sensitivity is increased by about 8% to 12% (for stainless steel collectors), depending on surface cleanliness and coverage by a hydrocarbon layer [10]. Due to the heating of the cathode it can be expected that the gas temperature in the gauge is higher than 293 K to 300 K by about 30 K [11], which reduces the sensitivity by about 10%. The two effects approximately cancel each other.

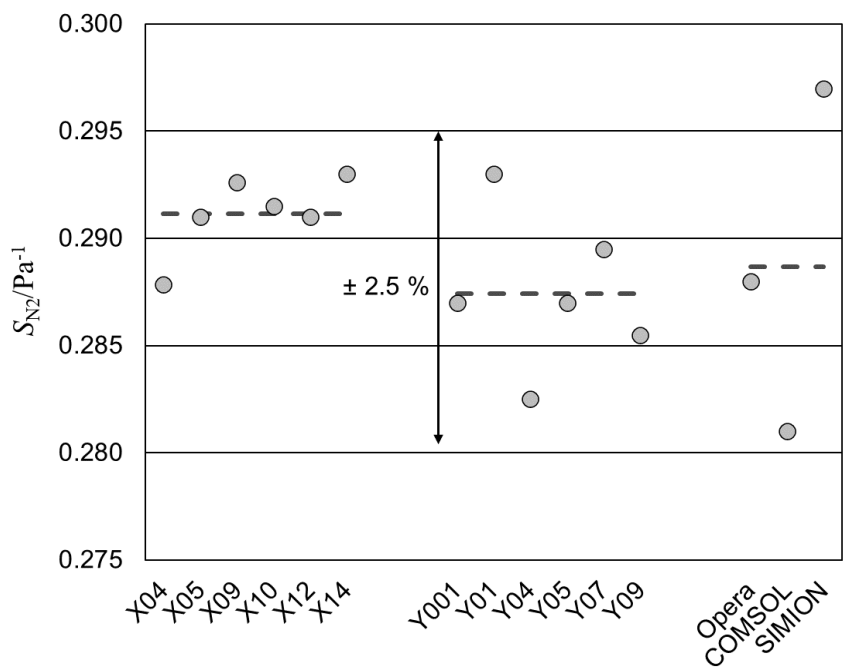

Figure 4 *Measured sensitivities for $N_2$ of six gauges from manufacturer X and five gauges from manufacturer Y as well as three calculated sensitivities from different simulation software packages*

The reproducibility of manufacturing can be quantified by the standard deviation of sensitivities. These data are collected in Table 1. The relative standard deviations were 0.63% and 1.39% for manufacturer X and Y, respectively, which shows excellent reproducibility of manufacturing.

The relative difference of $N_2$ sensitivities between the types of the two manufacturers is 1.26 %. It must be noted that the number of samples is still relatively small, and the sensitivities were measured in different National Laboratories, which means that the difference, the mean values and standard deviations of manufacturers X and Y include possible bias between the different measurement standards.

Table 1 Mean values and standard deviations of measured sensitivities for nitrogen of manufactured gauges by two manufacturers. The difference of the mean values is also given.

|  | $S_{N2}$ mean value [Pa$^{-1}$] | $u(S_{N2})$ stand.dev. [Pa$^{-1}$] | $u_{rel}(S_{N2})$ Rel. stand.dev. |
|---|---|---|---|
| Manufacturer X (6 gauges measured) | 0.2912 | 0.0018 | 0.63% |
| Manufacturer Y (5 gauges measured) | 0.2875 | 0.0040 | 1.39% |
|  | Difference of two manufacturers |  | relative difference |
|  | 0.0037 |  | 1.26% |

Another important characteristic of any measuring instrument is linearity. This was tested with nitrogen gas in the pressure range from $1\times10^{-6}$ Pa to $1\times10^{-2}$ Pa (Figure 5). All gauges tested showed a linearity well within $\pm 0.5$ % (Table 2).

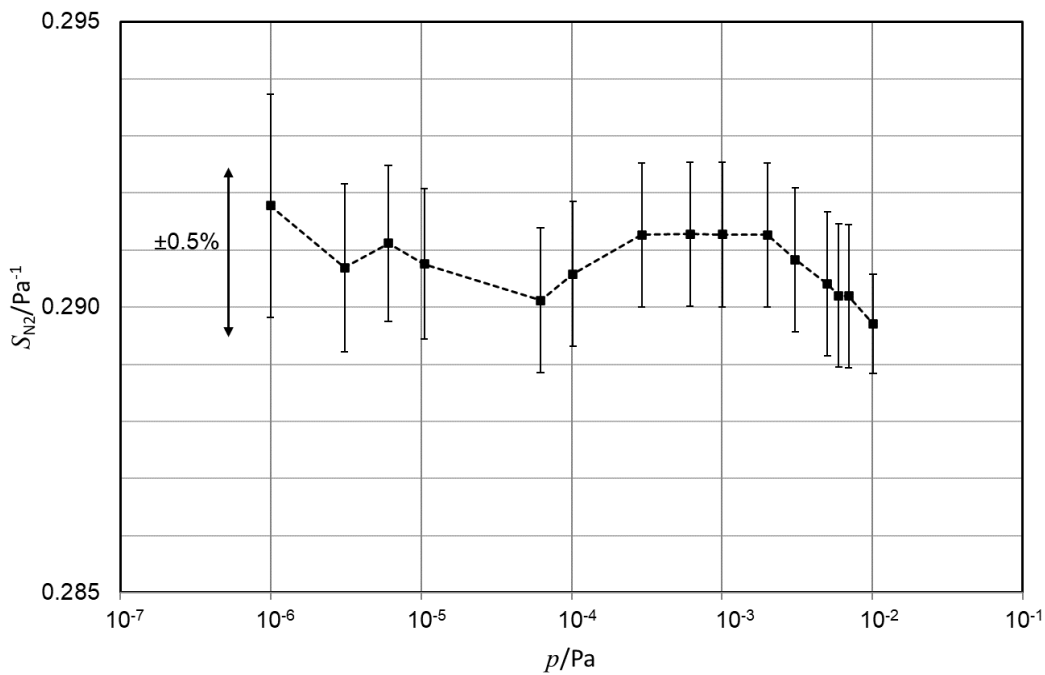

Figure 5 *Measured sensitivities with uncertainties for N$_2$ in dependence of pressure realised in the calibration standard of LNE. Gauge X10.*

We also checked if there is a depletion of molecules within the electron beam [5]. This would be revealed by a non-linearity of ion current versus electron current. In N$_2$ at $1\times10^{-2}$ Pa and up to 300 µA electron current no non-linearity was found.

The gauge also showed an excellent repeatability. Ten sensitivity measurements were taken at the same pressure within one hour. Between the measurements the system was evacuated to a base pressure below $1.5\times10^{-7}$ Pa. Relative standard deviations were 0.019% at a pressure of $1.0\times10^{-5}$ Pa (Table 2) and 0.044 % at $1.0\times10^{-4}$ Pa.

The resolution limit is mainly determined by the resolution of the current meter measuring the ion current and the stability of the residual current. So far, a resolution down to $2\cdot10^{-14}$ A was demonstrated at measured residual offset current of 1 pA. With an emission current of 30 µA and the nominal sensitivity for nitrogen of 0.29 $Pa^{-1}$, the pressure resolution limit is $2\cdot10^{-9}$ Pa.

Another important goal of the design was stability over time and stability after regular operations with the gauge like venting, bake-out, idling at residual pressure, exposure to other gases, cathode exchange, and transport. To quantify the reproducibility of sensitivity over time of individual gauges we define the quantity

$$\Delta = \frac{S_i - \bar{S}}{\bar{S}} \qquad (3)$$

as relative difference of sensitivity $S$ at measurement $i$ to the mean value of sensitivity $\bar{S}$ obtained for this gauge and gas.

Figure 6 shows $\Delta$ over a period of 5 days or more, for gauges Y04, X04 and X10, measured at three different laboratories. Sensitivities were measured for argon and nitrogen gas. Before the first measurements the gauges were baked at 150 ºC for a period of 48 h to 110 h, followed by an operation at an Ar pressure of 5 mPa ("conditioning", see Section 3). At PTB and LNE the duration of this conditioning was 1 h, and at IMT 2.5 h for gauge Y04 and 1.6 h for X04.

A few of the stability tests of the different gauges are described in the following.

For gauge Y04, after $N_2$ sensitivity measurements on Day 2, sensitivities for active gases $O_2$, CO and $CO_2$ at pressure $3\times10^{-4}$ Pa were measured, followed by operation at residual pressure below $3\times10^{-7}$ Pa for 16 h until the next $N_2$ sensitivity measurements on Day 3. Then, the gauge was left in operation at residual pressure for 5 days before the final measurement on Day 8.

Gauge X04, measured at IMT, received a second and third conditioning with a duration of 3 h at Ar pressure of 5 mPa before measurements on Day 4 and 5, respectively. On Day 7 sensitivities for the "active" gases $O_2$, CO and $CO_2$ at pressures of $3\times10^{-5}$ Pa (10 times lower pressures than with gauge Y04) were measured. The gauge was then left in operation at base pressure below $2\times10^{-7}$ Pa for 14 h until the final measurement on Day 8.

Gauge X10 at LNE was included in a study of repeated conditioning at an Ar pressure of 5 mPa. The duration of conditioning was 1h before measurements on Day 5 and 6, and 2h before measurements on Days 7 and 8. Then the gauge was exposed to atmosphere and baked at 150 ºC for three days. After bakeout, the sensitivity was first measured without Ar conditioning on Day 13. After that, two more sensitivity measurements were performed on Days 14 and 15 after Ar conditioning for 2 h on each day.

It can be seen that Δ after all these procedures was within ± 1% for the 3 gauges.

To test the robustness of the design after repeated venting and baking, an experiment with 5 repeated cycles was performed (Figure 7). Each cycle started with venting to atmosphere for 1 h, followed by a vacuum bakeout at 150˚C for 24 h and a cool-down phase to 30˚C. The gauge was put in operation at residual pressure for at least 12 h, after which it was conditioned in Ar for 1 h at 30 μA emission current. The pressure was $1\times10^{-2}$ Pa for runs No. 1, 4 and 5, and $1\times10^{-3}$ Pa for runs No. 2 and 3. Measurement of sensitivity for Ar and $N_2$ followed at a pressure of $1\times10^{-3}$ Pa. Values of sensitivity after these repeated venting and baking periods were within ± 1% (Figure 7).

Table 2 Summary of various experimental tests of the novel ionisation vacuum gauge.

| Test | Number of gauges | Relative change of sensitivity for nitrogen or standard deviation |
| --- | --- | --- |
| Non-linearity $10^{-6}$ Pa to $10^{-2}$ Pa | 11 | < 0.5% (variation) |
| Repeatability at $10^{-5}$ Pa after pump down, 10 measurements | 1 | 0.019 % (st.dev.) |
| Reproducibility after repeated venting and bake-out | 5 | < 1% (change) |
| Reproducibility after exposure to $O_2$, CO and $CO_2$ | 2 | < 1% (change) |
| Cathode exchange | 1 | 0.34% (change) |
| Simulated transport and drop-down test | 1 | 0.34% (change) |
| Real transport instability during comparison between four European laboratories | 2 | 0.57% (change after 4 months) 0.25 % (change after 7 months) |

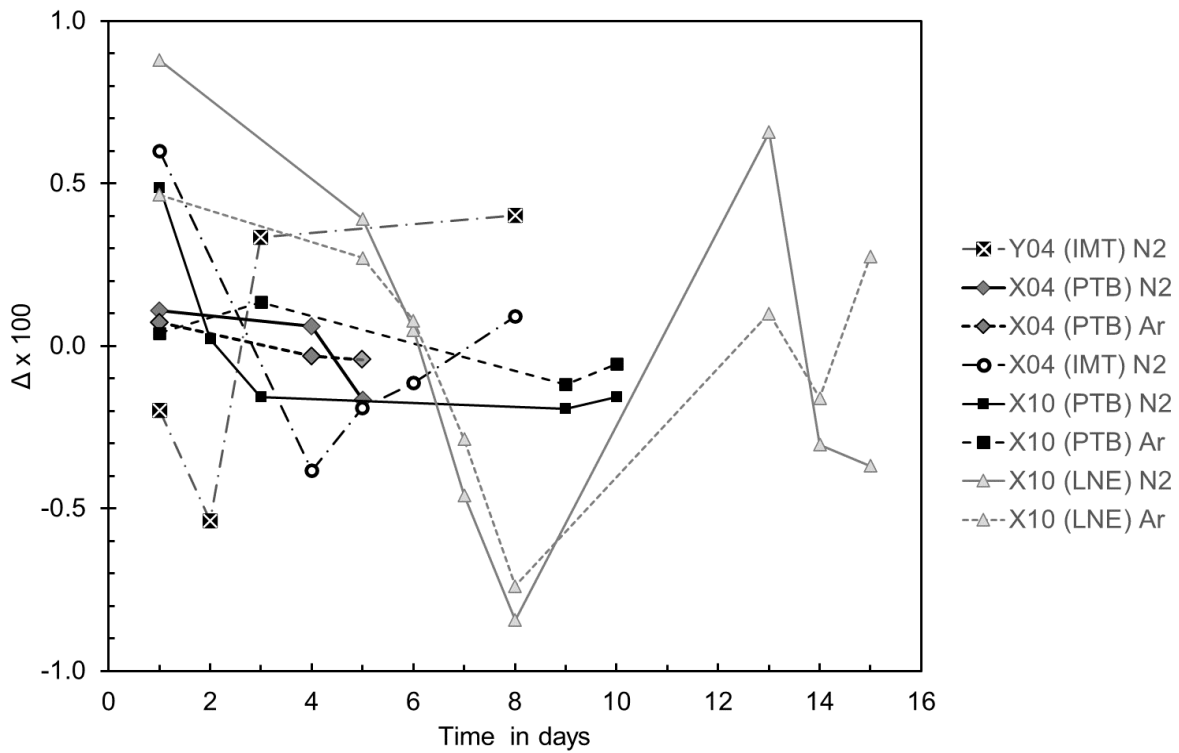

Figure 6 *Reproducibility of the sensitivity over 5 or more days for three selected gauges Y04, X04 and X10 (see text). Gauge X10 at LNE was vented to atmosphere and baked between measurements on Day 8 and Day 13.*

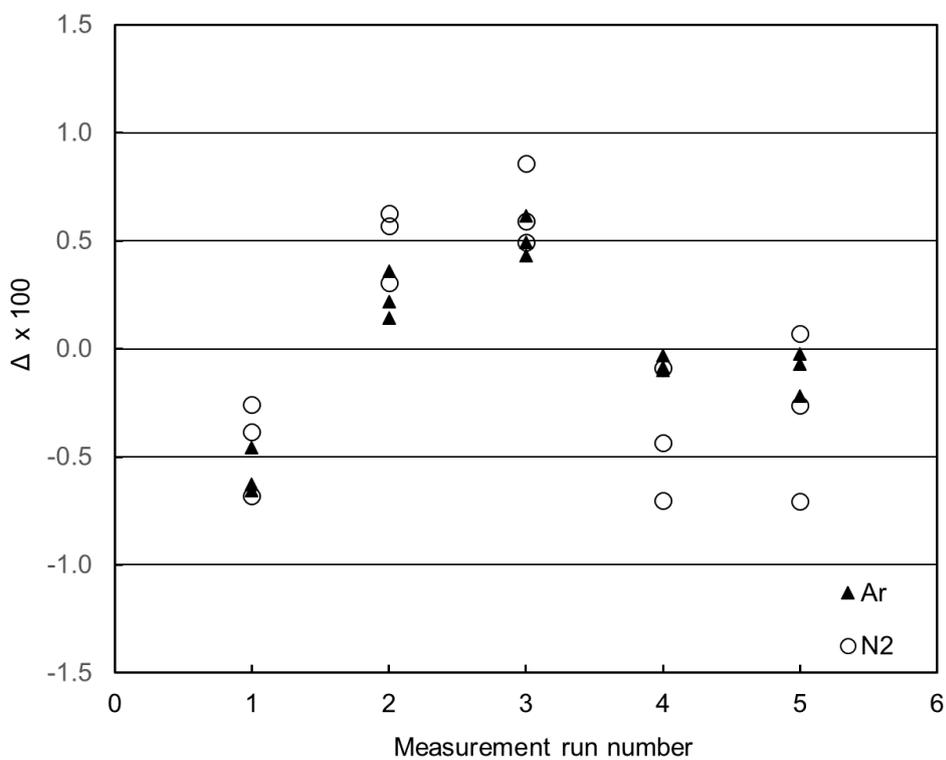

Figure 7 *Reproducibility of the sensitivity for Ar and $N_2$ after repeated venting and baking (gauge X014). For runs No. 1, 4 and 5, conditioning was done at a Ar pressure of $1 \times 10^{-2}$ Pa, and for runs No. 2 and 3 at $1 \times 10^{-3}$ Pa.*

For one of the prototype gauges (X012) we checked sensitivity after a replacement of the cathode by an identical one. The sensitivity for $N_2$ was 0.291 $Pa^{-1}$ for the original cathode and changed to 0.290 $Pa^{-1}$ after the cathode exchange which is within the measurement uncertainties. This means that a cathode exchange does not require re-calibration. The electron transmission changed by 0.34% only.

To check the stability after a transport, another gauge was packed in a cardboard box and exposed to a road transport in a car trailer for a total of 740 km, followed by an impact test. The box with the gauge was dropped 10 times from 1 m height (1 × corner, 3 × edge, 6 × face).

Measurements after the transport and drop-down tests with the same gauge showed a reduction of the electron transmission efficiency to 90 % from 95 % before the test. The shape of the transmission function over Wehnelt potential was typical for a displaced cathode. This displacement of the cathode, however, had little effect on the $N_2$ sensitivity. Its value was 0.2933 $Pa^{-1}$ before the test, and 0.2923 $Pa^{-1}$ after, so the relative change was -0.34% only (Table 2).

A second transport test was carried out as a laboratory comparison between PTB, CMI, IMT and LNE. Two gauges were selected as transfer standards in separate loops. For the first loop, the $N_2$ sensitivity measured at PTB at the beginning of the comparison was 0.2809 $Pa^{-1}$. When the gauge had returned to PTB after 4 months the sensitivity was remeasured to 0.2793 $Pa^{-1}$, corresponding to a relative change of -0.57%. For the second loop, the initial value was 0.2868 $Pa^{-1}$ and the final value (measured after 7 months) was 0.2861 $Pa^{-1}$, both measured at PTB, corresponding to a relative change of -0.25% (Table 2). These are excellent stabilities compared to transport instabilities of past comparisons [12], [13], where the changes after transport were in the range from 2% to 7%.

5. **Conclusions**

This new design of an ionisation vacuum gauge allows to predict its sensitivity by a careful guiding and shaping of the ionising electron beam: The Wehnelt cylinder around the disk emitter controls the emission point and the direction of electron trajectories into the anode, the ion collector focuses the electron beam into the anode exit, a deflector directs the electrons in a U-turn onto a Faraday cup, so that soft X-ray photons cannot hit the ion collector and secondary electrons cannot enter the ionisation volume.

The new gauge, designed for a measurement range from $10^{-6}$ Pa to $10^{-2}$ Pa, was tested by experiments and calibrations with fundamental standards for high vacuum at National Metrology Institutes. The predicted sensitivity was accurately confirmed by calibrations. The spread of sensitivity of different individual gauges from two different manufacturers was below ±2.5 %. The non-linearity is below 1%. Transport, venting and bake-out tests showed a stability unmatched by previous ionisation vacuum gauge types. A cathode exchange does not significantly change the sensitivity.

The reproducibility and robustness of the gauge is mainly caused by its rigid and robust structure (addressing Problem 2), while present BA gauges are made of wires and meshes. Such a structure is difficult to manufacture in a reproducible manner and are prone of changes after shocks and vibrations.

The excellent characteristics make this gauge a suitable candidate for the description of an ISO Technical Specification, so that it can be produced by any experienced manufacturer. The fact that the spread of individual sensitivities between the gauges is so small may also have a commercial impact, because manufacturers need not to calibrate the gauge before delivery and users can exchange the gauge without re-calibration and re-adjustment of their process.

This new gauge may also serve as a new accurate reference gauge for high vacuum and as a transfer standard for comparisons between measurement standards of calibration and National Laboratories. It will also improve pump speed measurements according to the ISO 21360 series, since relative sensitivity factors will be accurately known.

**Acknowledgements**

The authors are glad to have written this paper in honour of the anniversary of John Colligon as editor, who supported in this role vacuum science and metrology for 40 years. The authors are also very grateful to M. Wüest and F. Scuderi from the INFICON AG, and M. Granovskij and Chris Gruber from the VACOM GmbH company for the professional manufacturing of the prototype gauges. INFICON also supported us with simulations by the COMSOL package. Authors from PTB are grateful to Dietmar Drung and Martin Götz for the development and calibration of the ultra-stable low-noise current amplifier (ULCA). This project 16NRM05 Ion gauge has received funding from the EMPIR programme co-financed by the Participating States and from the European Union's Horizon 2020 research and innovation programme, and the Portuguese National Funding Agency for Science, Research and Technology in the framework of the project UID/FIS/00068/2019.